# A FEL Based on a Superlattice

I.V. Dovgan*

Department of Physics, Moscow State Pedagogical University, Moscow 119992, Russia.

*dovganirv@gmail.com

**Abstract**

The motion and photon emission of electrons in a superlattice may be described as in an undulator. Therefore, there is a close analogy between ballistic electrons in a superlattice and electrons in a free electron laser (FEL). Touching upon this analogy the intensity of photon emission in the IR region and the gain are calculated. It is shown that the amplification can be significant, reaching tens of percent.

**Introduction**

Creation of compact inexpensive sources of radiation operating efficiently in visible, UV, or soft X-ray domains is one of the most important directions in the development and investigation of Free-Electron Lasers. A short-wavelength radiation can be generated by a FEL using either a high-energy (multy-GeV) electron beam or a short period undulator. One of the ideas often mentioned and discussed in the context FEL theory is that of a FEL using Media with a Periodically Modulated Refractive Index (MPMRI). MPMRI can be considered as a kind of a volume diffraction grating. The following two types of MPMRI could been realized: (1) a gas-plasma medium with periodically varied density or degree of ionization [1] and references therein and (2) a spatially periodical solid-state superlattice-like (SL) structure, which can be composed, e.g., of a series of layers of different materials with different refractive indices [1] and references therein. The experimental feasibility of a transition radiation FEL was shown recently [2]. The macroscopic channeling of electrons in an intensive standing light wave is considered in [3,4].

To obtain IR radiation with variable frequency we propose to use a SL where the currents are injected and passed through without collisions with phonons. As the mean free path for collisions is about $2 \cdot 10^5 cm$, the length of the SL has to be smaller than this. We restrict our consideration to a SL of the GaAs/GaAsAl type. Other types may work also, for example, InGaAs/AlGaAs. The injection of carriers in the case of GaAs/GaAlAs SL may be performed as suggested in Ref. [5]. The injection system consists of a heavily doped GaAs layer (thickness 2000 A, carrier density $n = 10^{18} cm^{-3}$) and a barrier of AlGaAs (thickness 100 A, undoped). Between the emitter and the SL a potential $V_E$ is applied [6,7].

**Two Observing Effects**

With the help of the potential $V_E$ the barrier is lowered and carriers from the emitter are able to pass ballistically through the SL. In such system two effects may be observed:

1) Spontaneous emission of IR photons by the carriers due to their interaction with the SL periodic potential. Carriers with a kinetic energy larger than the potential amplitude $U_0$ of the SL begin to oscillate in the direction of their motion. Therefore the photons emitted are directed mainly perpendicular to the axis of the SL.

Let us consider a beam of electrons injected along the SL axis X. The SL potential is chosen in the form $U(x) = U_0 \cos qx$, where $q$ is the SL wave vector. Then the spectral intensity of dipole radiation by one electron has following form

$$\frac{dI}{d\Omega d\omega} = \frac{e^2 v^2}{64 c^3} \sin^2 \theta \left(\frac{eU_0}{E}\right)^2 \frac{\sin^2 u}{u^2} \varepsilon_0 n^2, \qquad (1)$$

where v is the velocity of the ballistic electron, $E = m^* v^2 / 2$, $u = n((\omega - qv)/\omega)\pi$, $\theta$ is the angle between x and the radiation direction, $\varepsilon_0$ is the dielectric constant of the SL, and $\frac{\sin^2 u}{u^2}$ is a diffraction factor.

Relationship (1) has a maximum when u approaches zero. By this is meant that the frequency of spontaneously emitted photons by ballistic electrons in the SL may be obtained from the expression

$$\omega = qv \qquad (2)$$

As the value of the velocity depends on the injection potential, variation of this potential leads to a variation of $\omega$.

Using Eq. (1) we can obtain the total intensity $I$ emitted by the electron:

$$I = \frac{\pi}{3} \frac{e^2}{\hbar c} \omega \left(\frac{v}{c}\right)^2 \left(\frac{eU_0}{E}\right)^2 \frac{\varepsilon_0}{16}. \qquad (3)$$

In reality the number of electrons, N, which may emit photons in the SL depends on the current density $j_0$ and the SL parameters

$$N = \frac{j_0 S l_0}{ev} \qquad (4)$$

S is the cross-section of the SL, and $l_0$ its length. The total intensity of emission from an electron beam is equal to

$$I_N = IN = \frac{j_0 S l_0}{ev} \frac{e^2}{\hbar c} \omega \left(\frac{v}{c}\right)^2 \left(\frac{eU_0}{E}\right)^2. \qquad (5)$$

The value of $I_N$ may be estimated for some reasonably chosen parameters of the SL and the injected electron current [5]( $S \sim 10^{-2} cm^2$, $j_0 \sim 10^2 A/cm^2$, $l_0 \sim 10^{-5} cm$, $v = 10^8 cm/s$, d = $2n r/q$ = 200 A, $\omega = 3 \times 10^{14} s^{-1}$ ($\hbar\omega = 0.2 eV$), $eU_0 / E \sim 1$, $\varepsilon_0 = 12.5$) leading to $I_N = 10^{14}$ photons/s = $= 7 \times 10^{14} W$.

2) Gain measured in the direction perpendicular to the axis of the SL. (The same direction is the direction of spontaneous radiation.) Now we will estimate the gain. The Hamiltonian describing the motion of an electron of energy E in the field

$$E = E_1 \cos(\omega t - kz) \tag{6}$$

may be written in the form [8-16]

$$\hat{H} = \hat{U}(x) + \hat{H}' + \hat{H}_0, \tag{7}$$

where $\hat{H}_0$ is the kinetic energy operator and $\hat{H}'$ is the electron-wave interaction operator $\hat{H}' = -(e\varepsilon \lambdabar/(m^*c))\hat{p}$, where $\lambdabar$ R is the wavelength of the electromagnetic field.

The electromagnetic wave (6) is polarized along the SL axis and propagates in a perpendicular direction to the SL axis.

Without the electromagnetic wave, the motion of an electron in the quasiclassical approximation may be written in the form

$$\psi_0 = exp\left(\frac{i}{\hbar}\int_0^x (p^2 - 2m^*eU_0 \cos qx')^{1/2} dx' - \frac{iEt}{\hbar}\right), \tag{8}$$

where p is the electron momentum, $E > eU_0$. In the presence of the wave an electron may emit or absorb a light quantum. Therefore in a first approximation the wave function will be of the form

$$\psi = \psi_0 + a_1(x)\psi_1 \exp(ikz - i\omega t) + a_{-1}(x)\psi_{-1}\exp(-ikz + i\omega t). \tag{9}$$

Here $\psi_1$ and $\psi_{-1}$ follow from the expression for $\psi_0$ by replacing the initial momentum by $p_1$ or $p_{-1}$ respectively, with $p_{\pm 1} = (p^2 \pm 2m^*\hbar\omega)^{1/2}$. $a_{\pm 1}$ are the amplitudes of absorption and stimulated emission, respectively ($a_{\pm 1} \ll 1$). Using perturbation theory we may obtain the amplitudes in the form

$$a_{\pm 1} = \mp\frac{e\varepsilon_0}{4\hbar\omega}\frac{eU_0}{E}\exp\left(\frac{i}{\hbar}(p - p_{\pm 1} \pm q)\frac{x}{2}\right) \times \frac{\sin(p - p_{\pm 1} \pm q)\frac{x}{2}}{p - p_{\pm 1} \pm q}. \tag{10}$$

The amplitudes have maximum values if the condition $p_{\pm 1} - p = \pm q$ is fulfilled. Within the limit $E \gg \hbar\omega$ this condition can be written as

$$\frac{\omega}{v} = q \pm \frac{p}{8\hbar}\left(\frac{\hbar\omega}{E}\right)^2. \tag{11}$$

This is consistent with Eq. (2) because

$$\frac{p}{8\hbar}\left(\frac{\hbar\omega}{E}\right)^2 \ll q.$$

Using the expression for the perturbed wave function (9). the gain may be calculated. The gain is proportional to the difference between the emission and absorption probabilities $W_0 = W_e - W_a$.
In our case we have:

$$W_e = |a_{-1}|^2, \quad W_a = |a_{+1}|^2. \tag{12}$$

As a result the gain is

$$G = 8\pi \frac{j_0}{e} W_0 \frac{\hbar\omega}{c\varepsilon_1^2} \sqrt{\varepsilon_0}. \tag{13}$$

If $\Delta E / E < 1/(\pi n)$, where n is the number of the SL periods, and $\Delta E$ is the energy spread of the injected electrons, then [6,17-24]

$$G = \frac{\pi}{16} \frac{j_0}{e} r_0 \frac{l_0 t^2 c^2}{\hbar v} \left(\frac{eU_0}{E}\right)^2 \times \frac{d}{du}\left(\frac{\sin^2 u}{u^2}\right) \sqrt{\varepsilon_0}, \tag{14}$$

where $r_0$ is the classical radius of the electron, and t is the interaction time.

The value of t is determined by the time for light to cross in the direction perpendicular to the SL axis: $t = l_\perp \sqrt{\varepsilon_0}/c$. Estimating the gain per pass using the same SL parameters and current density as before, we obtain G = 0.1%.

To increase the gain up to a reasonable value ( = 10%) the current density has to be increased to $10^4 A/cm^2$. This current density seems feasible in a pulsed regime. If $\Delta E / E > 1/(\pi n)$, the expression for G has the following form:

$$G = \pi^2 \frac{j_0}{e} r_0 t\hbar (eU_0)^2 \frac{df}{dE}\Big|_{E=E_0} \sqrt{\varepsilon_0}, \tag{15}$$

where f is the distribution function of the electron beam. and $E_0 = m^*\omega^2/(2q^2)$. If we take $df/dE \sim 1/(\Delta E)^2$, $U_0/\Delta E \sim 10$, and $j_0 = 10^4 A/cm^2$, then G = 10%.

## Conclusion

The frequency of the light amplified may be easily changed by variation of the carrier velocity. The gain may be increased if several SL are combined in the direction of light propagation like in the case of multilayer system transition radiation.

## References

1. G.A. Amatuni, A.S. Gevorkyan, S.G. Gevorkian, A.A. Hakobyan, K.B. Oganesyan, V. A. Saakyan, and E.M. Sarkisyan, Laser Physics, **18** 608 (2008).
2. H.C. Lihn, P. Kung., C. Settakron, H. Wiedemann, D. Bocek, M. Hernandez, Phys. Rev. Lett. **76**, 4163 (1996).
3. Fedorov M.V., Oganesyan K.B., Prokhorov A.M., Appl. Phys. Lett., **53**, 353 (1988).